\documentclass[draft,wrr]{agu2001}
\newcommand{\W}{14cm}
\usepackage{graphicx,amsmath,amssymb}

\authorrunninghead{Auradou et al.}
\titlerunninghead{Permeability anisotropy induced by a shear displacement}

\authoraddr{H. Auradou,Laboratoire Fluide, Automatique et Syst{\`e}mes
Thermiques, UMR No. 7608, CNRS, Universit{\'e} Paris 6 and 11,
B{\^a}timent 502, Universit{\'e} Paris Sud, 91405 Orsay Cedex, France.
(auradou@fast.u-psud.fr)}
\begin{document}

\title{Permeability anisotropy induced by the shear displacement of rough fracture walls.}

\authors{H. Auradou, \altaffilmark{1}
G. Drazer, \altaffilmark{2}
J.P. Hulin, \altaffilmark{1}
and J. Koplik \altaffilmark{2}}
\altaffiltext{1}
{Laboratoire Fluide, Automatique et Syst{\`e}mes Thermiques,
UMR No. 7608, CNRS, Universit{\'e} Paris 6 and 11, B{\^a}timent 502,
Universit{\'e} Paris Sud, 91405 Orsay Cedex, France.
}
\altaffiltext{2}
{Benjamin Levich Institute and Department of Physics, City College of the
City University of New York, New York, NY 10031.
}

\begin{abstract}
The permeability anisotropy that results from a shear displacement ${\vec u}$ between the 
complementary self-affine walls of a rough fracture is investigated.
Experiments in which a dyed fluid displaces a transparent one as it is radially injected into 
a transparent fracture exhibit a clear anisotropy in the presence of shear displacements, 
and allow us to estimate the ratio of the permeabilities for flows parallel and perpendicular to ${\vec u}$.
A simple model which accounts for the development of channels perpendicular to ${\vec u}$ qualitatively 
explains these results, and predicts  a permeability decreasing (increasing) linearly with the 
variance of the aperture field  for flow parallel (perpendicular) to the shear displacement. 
These predictions are then compared to the results of numerical simulations performed using a 
lattice-Boltzmann technique and to the anisotropies measured in displacement experiments.
\end{abstract}

\begin{article}
\section{Introduction}
Subsurface fluid flow in low-permeability geological formations occurs primarily through networks of 
fractures \cite[]{nas,adler99,sahimi95}. In order to model such systems, one needs  to understand
flow in single fractures  which are the  building blocks of the network.
Among the many parameters which influence fluid motion at this scale, we focus here on the geometry 
of the fractures, specifically on the correlations between the roughness of the surfaces of the two 
opposite walls. 
A shear displacement of the latter alters these correlations and previous laboratory measurements 
demonstrated that it also affects the permeability, through variations of the local aperture of the 
fracture due to the wall roughness \cite[]{PlouraboueJCH00,OlssonB93}.

In related experiments, a higher permeability was measured for flow perpendicular to the shear 
displacement than parallel to it \cite[]{GentierLAR97,YeoDZ98}. Numerical studies found a similar 
behavior on explicitly anisotropic rough surfaces generated numerically \cite[]{ThompsonB91};
an indication of such anisotropic permeability was also observed in a previous work on self-affine 
fracture \cite[]{Drazerk02}.
The purpose of the present paper is to combine experiments, systematic numerical simulations, and 
analytic arguments to understand the permeability anisotropies that are induced by shear
displacements of rough fracture walls.

A key geometrical feature of the fracture surfaces studied in this work is their statistical scale 
invariance known as {\it self-affinity} \cite[]{mandelbrot83,feder88}.
The analysis of the surfaces of both fault and fresh fractures has established that their roughness 
cannot be described by a finite set of typical wavelength values, but instead is self-affine.
A self-affine surface is one that is {\it statistically} unchanged under the scaling relation:
$x \rightarrow \lambda x$, $y  \rightarrow \lambda y$ and $h(x,y) \rightarrow \lambda^\zeta\ h(x,y)$.
($(x,y)$ are coordinates in the mean surface plane, $h(x,y)$ is the local height of the surface, 
and $\zeta$ is known as the Hurst exponent.) 
The self-affine character of the fracture surfaces has been observed, over a significant range of 
length scales, 
in a broad variety of materials and for both natural and man-made fractures \cite[]{Bouchaud03}.
In the case of fractured rocks, such as granite or basalt rocks, the Hurst exponent is found to be 
$\zeta\approx 0.8\pm0.05$ over a broad range of length scales \cite[]{Bouchaud03}, although values
near $0.5$ are found in other rocks \cite[]{BoffaEAP98}.

The key consequence of self-affinity in our context is the following scaling law for the 
variance of the surface heights:
\begin{equation}
\langle [h(\vec{x}+\vec{\lambda})- h(\vec{x})]^2\rangle \sim
|\vec{\lambda}|^{2\zeta},
\label{eq:hh}
\end{equation}
which implies long-range correlations in the height-to-height fluctuations, 
and should be contrasted with the fluctuations in surface height that would arise 
from a random field with a characteristic correlation length, which ultimately 
leads to a linear growth of the variance \footnote{There is a direct analogy between 
the variance in the surface heights and the random walk problem. A standard random walk
corresponds to a Hurst exponent $\zeta=0.5$ and a linear growth of the
variance, whereas a Hurst exponent $\zeta=0.8$ corresponds to an anomalous hipper-diffusive
behavior \cite[]{feder88}.}. It will be seen that these correlations strongly affect the 
spatial aperture distribution of the rough fractures studied in the present work, and 
therefore, the flow inside them.

In both experiments and numerical studies, we shall consider model fractures, made of two complementary rough 
self-affine surfaces: two perfectly matching surfaces separated by a distance $a_0$ in the direction normal 
to their common mean plane and, in general, subject to a shear displacement ${\vec u}$ parallel to this mean 
plane. In all cases, the gap between surfaces is large enough so that the two walls do not touch, 
thereby avoiding the effects of shear-induced dilation of the fracture gap on its permeability.

The experiments were performed using transparent models fabricated with moldings of actual rocks. 
In the experiments, a dye is injected radially from a point of the fracture and the anisotropy in
the permeability is computed from that of the injection patterns. 
Only the ratio of the highest and lowest permeabilities can be determined in this way, but not 
their absolute values. These experiments are therefore appropriately complemented by numerical 
simulations which, in addition to the permeability, provide the complete flow field. 
The lattice Boltzmann technique has been used since it is convenient for determining the velocity 
field in these complex geometries \cite[]{Drazerk02,kimLD03}.

The paper is organized as follows: in Sec.~\ref{subsec:sec1}, the laboratory procedure used to obtain the 
transparent cast of the fractured rock is described as well as the characterization of the surface roughness.
The experimental set-up and the image analysis methods are described in Sec.~\ref{subsec:sec2}. 
In order to explain the experimental results described in Sec.~\ref{sec:sec3}, a theoretical analysis of 
the relation between the aperture field and the observed anisotropy in the permeability is developed in 
Sec.  \ref{sec:sec4}. The scaling laws obtained in this way are then compared to numerical simulations 
in Secs.\ \ref{sec:sec5} and \ref{sec:sec6}. The same approach is finally applied to the experimental 
data in Sec.~\ref{sec:sec7}.

\section{Experimental set-up and procedure}
\label{sec:sec1}

\subsection{Model fracture and flow set-up}
\label{subsec:sec1}
The experiments were performed on transparent moldings of fractured granite blocks extracted from a quarry in 
Lanhelin (Britanny). A tensile crack was produced by a compressive force applied on two opposite sharp edges 
carved in the middle of two facing sides (the initial block size is $25\times 25 \times 40 \ cm^3$).
Then, a silicon rubber molding of one of the fracture surfaces is made, and after that used as a template to 
fabricate transparent epoxy casts of the original surface. 
A second silicon rubber mold of the epoxy molding is then produced to obtain a perfectly matching surface 
(dimensional deformations during polymerization are much smaller for the silicon than for the epoxy).  
This method is widely used and reproduction problems are negligible for the present application
\cite[]{AuradouHR01,YeoDZ98,IsakovOTG01,kostakisHH03}.
The complementary epoxy and rubber surfaces are the walls of the model fracture used in the present work.

\begin{figure}
\includegraphics[width=\W]{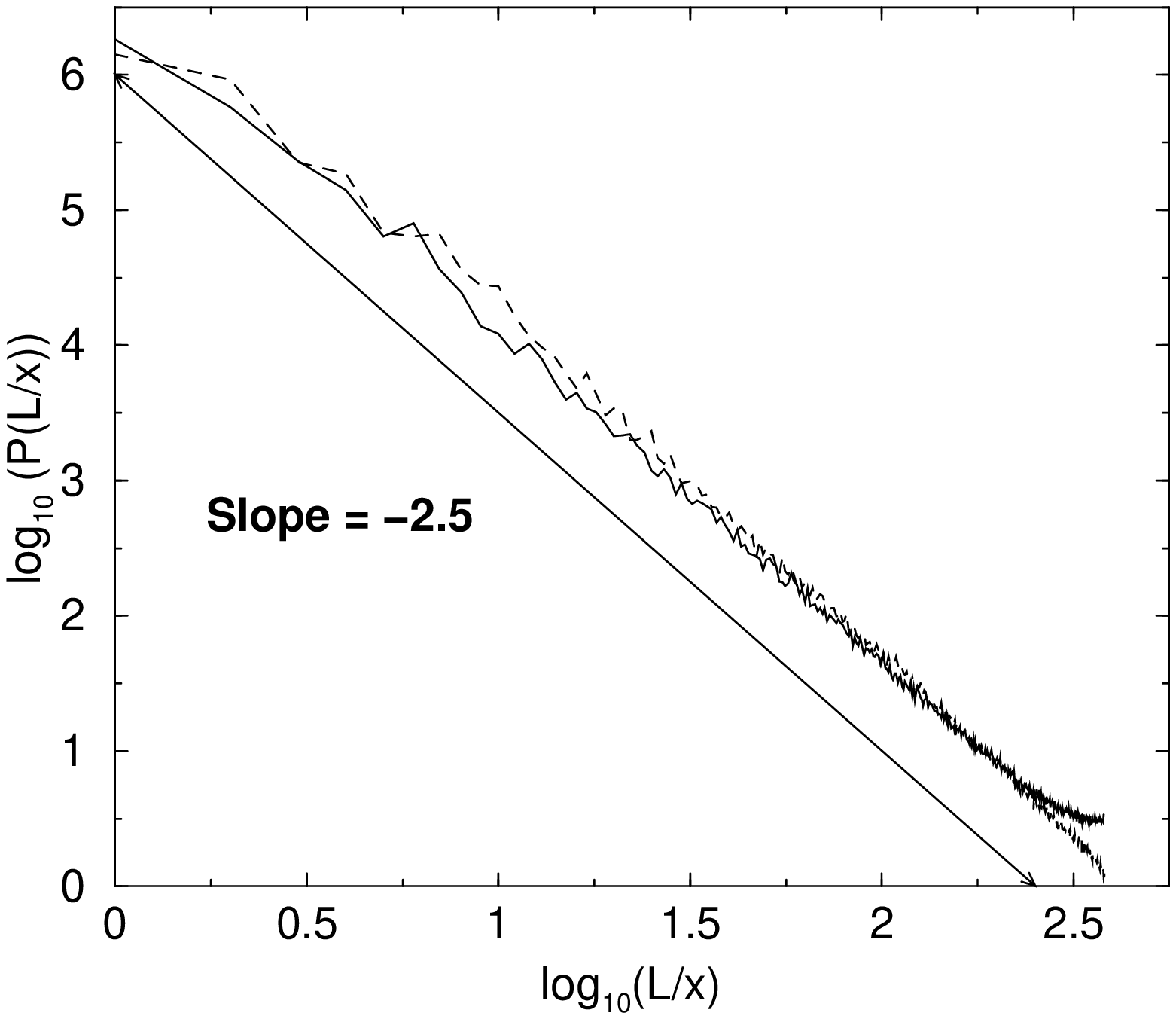}
\caption{Power spectral density of the experimental fracture
surfaces. The spatial frequencies are $f=L/(n\Delta x)$, for $n=1,2, \dots, L/\Delta x$,
where $L=20cm$ is the fracture length and $\Delta x=250\mu m$ is the resolution length
along the profile. The solid and dashed lines correspond to profiles obtained normal and
parallel to the direction of crack propagation.
The latter two curves are shifted vertically by $0.75$ for visual clarity.}
\label{fig1}
\end{figure}

Prior to the experiments, $2D$ maps of the epoxy surfaces are obtained by means of a mechanical profilometer 
($770\times760$) points with a $250\mu m$ grid spacing and a $10 \mu m$ vertical resolution ~\cite[]{BoffaEAP98}.
The differences in height between the highest and the lowest points of the surface is $\simeq 17\ mm$ and the 
root mean square deviations of the surface heights is $\simeq 3.5\ mm$.
As discussed in the introduction, {\it self-affine} surfaces have a roughness which does not contain any 
intrinsic wavelength, and it can be shown that the power spectral density $P(f)$ of the heights, for a 
linear profile drawn on the surface, should verify the following scaling law with the spatial frequency $f$ :
\begin{equation}
P(f) \sim A \times f^{-1-2\zeta}.
\label{eq:spec}
\end{equation}
Such spectral densities $P(f)$ obtained for the surface samples are plotted in  Fig. \ref{fig1}, 
and follow indeed a power law as a function of $f$.  The computed spectral densities showed no dependence,
within experimental error, on the orientation of the profile used to computed them, which implies that the
fractures are isotropic in their mean plane.
By averaging spectra obtained for several orientations of the profiles, one obtains $\zeta  = 0.75 \pm 0.05$, 
in agreement with previous studies of granite fractures~\cite[]{BoffaEAP98,MeheustS00,AuradouHR01}.
As indicated by the solid line with arrows at both ends in Figure \ref{fig1}, $P(f)$ follows this power law
over approximately two decades : the corresponding length scales range from $50\ mm$ down to $0.75\ mm$.  
The lower limit $f_m$ of this range is set by the resolution used in the present map; previous measurements have 
shown that the self-affine description remains valid at lower length scales, of the order of $0.1 mm$ or 
less~\cite[]{BoffaEAP98}.
The upper limit $f_M$ arises from the reduced number of wavelengths which describe the surface topography at 
large scales: $f_M$ is thus related to the finite size of the sample, which has a large-scale
topography dominated by a few large undulations. This statistical  effect usually becomes apparent at length 
scales greater than  one fourth of the system size \cite[]{MeheustS00}.

\subsection{Experimental procedure and image analysis}
\label{subsec:sec2}

The objective of the experiments is to analyse the flow inside an open fracture by monitoring the radial spreading of 
an injected dye. Using a radial injection geometry allows to estimate variations of the permeability of the fracture 
with the flow direction from  asymmetries in the  shape of the invaded zone.
\begin{figure}
\includegraphics[width=\W]{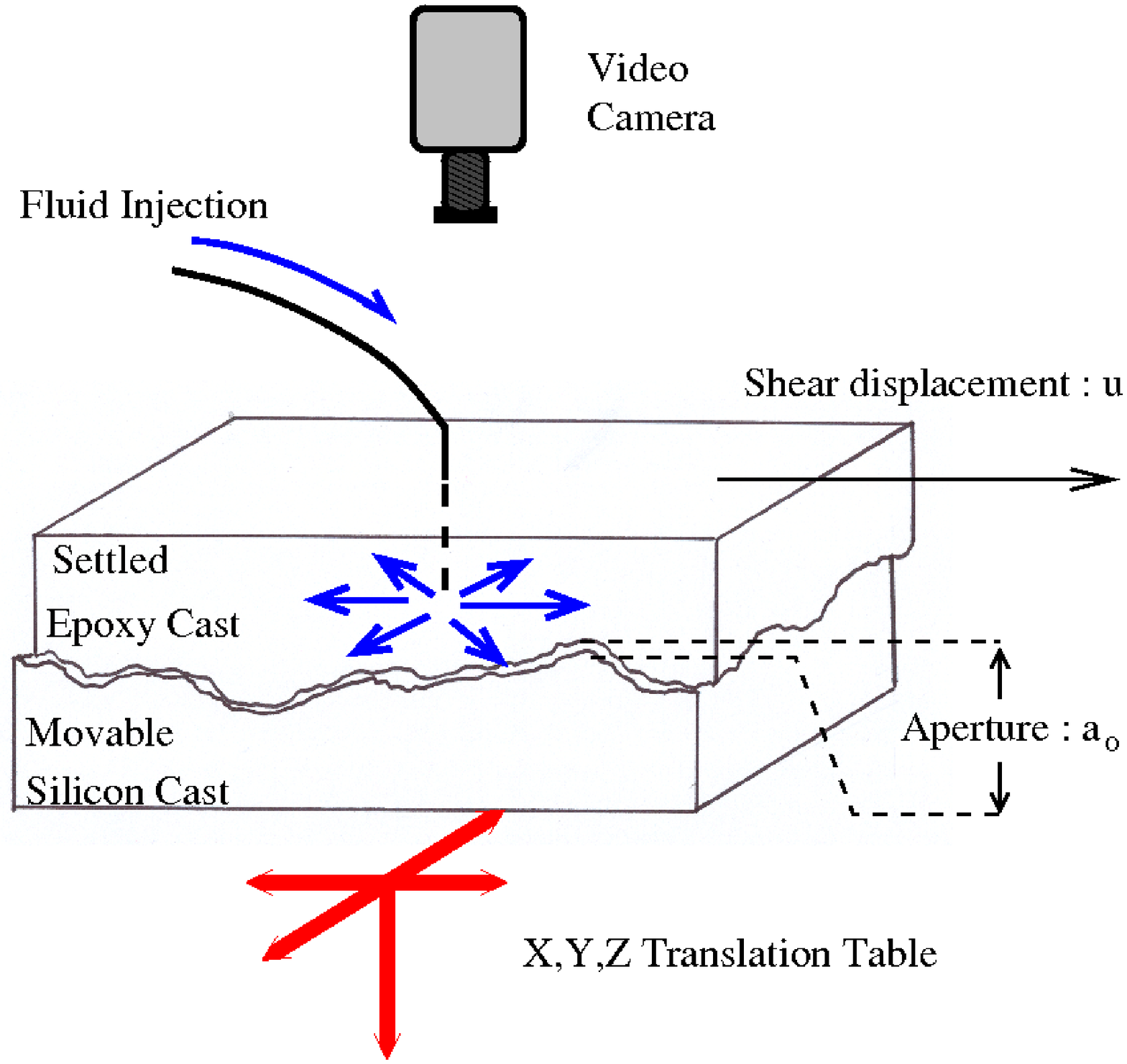}
\caption{Schematic view of the experimental set-up. The top wall is a transparent epoxy cast and has a fixed 
position. The bottom wall is a silicone replica of the top wall and can be moved down and laterally. The dyed 
fluid is injected in a radial, divergent configuration and it displaces a non-dyed fluid which initially 
saturates the fracture gap. The flow process is visualized with a camera set above the fracture. 
The two controlled lengths are indicated: $a_0$, the mean vertical separation between walls, and $\vec{u}$, 
the shear-displacement vector.}
\label{fig2}
\end{figure}
The two complementary surfaces are initially brought  into  full contact, with their mean plane  $(x,y)$ horizontal. 
The top (transparent) surface, see Fig.~\ref{fig2} for details, is kept fixed thereafter and
the lower one can be translated while remaining precisely parallel to its initial orientation.
Specifically, it is first displaced downwards ($-z$ direction) by a distance $a_0$ to open  the fracture and 
then laterally in the  $(x,y)$ plane to the required location. This lateral displacement  ${\vec u}$ is small 
enough so that the two surfaces do not come in contact.

The fracture model is then placed inside a leak tight basin which is filled with transparent fluid in order to 
saturate the model prior to the experiments.
Transparent fluid is then injected through a hole in the top surface and flows radially outwards. 
Once a steady flow is reached, a dyed fluid is injected and pictures of the invaded region are taken at constant 
time intervals using a high resolution cooled 12 bits digital camera.  
Each pixel corresponds to a $250\mu m \times 250\mu m$ area.
The fluids used are water-glycerol solutions with $10\%$  glycerol.  $0.2$-$0.3\ g/l$ of nigrosine are added to 
the dyed solution and the density of the two fluids is matched by adding an equivalent amount of $NaCl$ to the
transparent solution.  
In order to avoid density differences due to temperature variations the solutions are stored at the same 
temperature.  In all experiments density differences between the two fluids were always smaller 
than $25\ 10^{-4}\ g/cm^3$.

\begin{figure}
\includegraphics[width=\W]{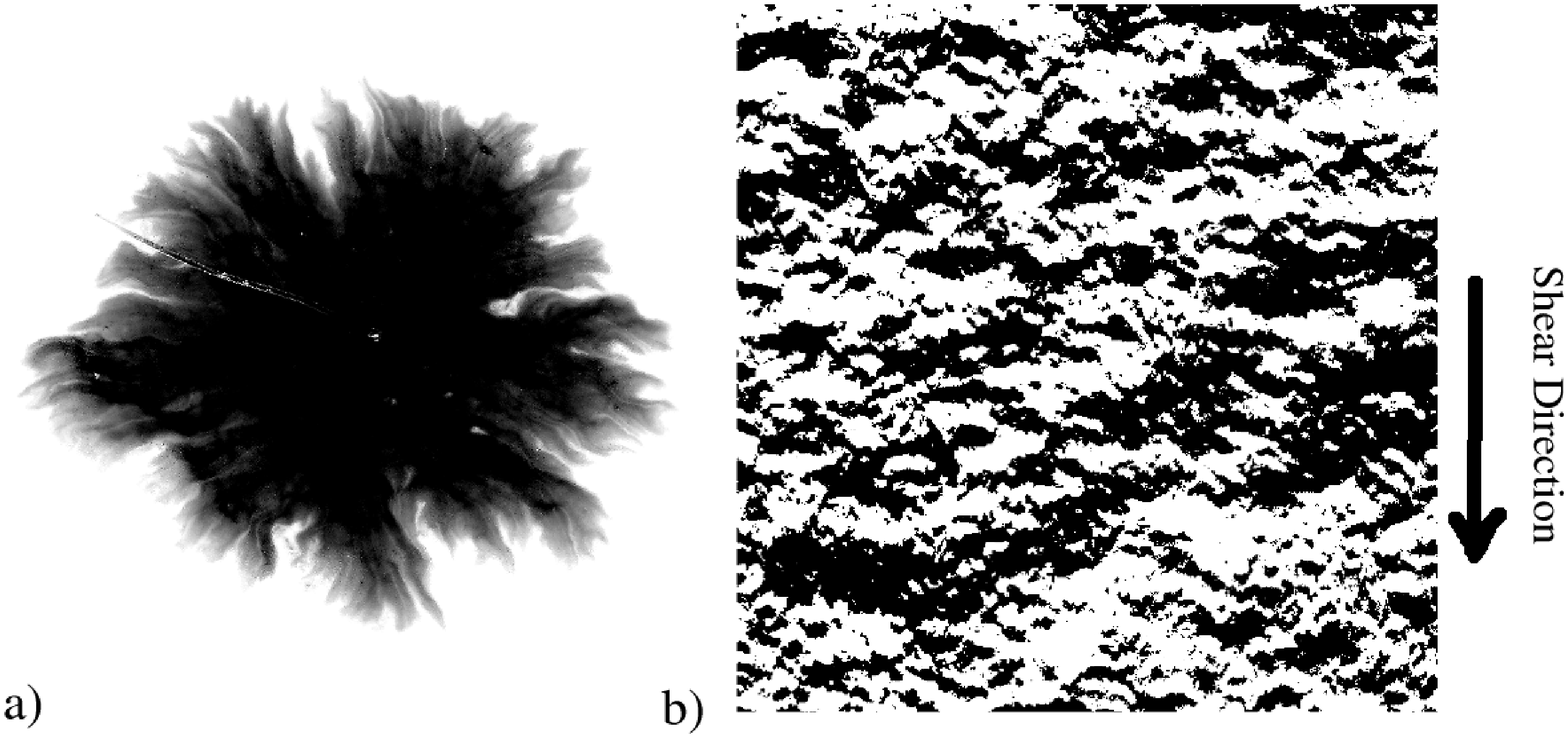}
\caption{Example of the relation between flow anisotropy and aperture field in a sheared fracture, 
with $a_0=1.25\ mm$ and $u_y = 2.0\ mm$ (the direction of the shear displacement is indicated by the
arrow). The normalized aperture variance is $S^2= 0.133$. 
 $(a)$ Concentration map.  $(b)$ Aperture field computed using Eq. \ref{eq:aper}, where white
corresponds to a local aperture $a(x,y) < a_0$ and  black to $a(x,y)>a_0$.}
\label{fig3}
\end{figure}

The dye concentration map is determined for each image obtained during the experiment by assuming the validity of 
Beer-Lambert's law. In this  case, the light intensity $I(x,y)$ at a given point $(x,y)$ of the image should 
vary exponentially both  with the local aperture $a(x,y)$ and the local concentration $c(x,y)$ with:
\begin{equation}
I(x,y)=I_0(x,y) exp\left(-2\ \mu\ c(x,y) a(x,y)\right),
\label{eq:bl}
\end{equation}
in which  $I_0(x,y)$ is the light intensity for transparent fluid and $\mu$ is the extinction coefficient.
The validity of this relation has been verified by opening the fracture without introducing a lateral shift so that
the local aperture $a(x,y)$ is constant and equal to the vertical displacement $\Delta z$.
The corresponding  light intensity $I(x,y)$ was then determined for different dye concentrations $c_0$ and different 
mean apertures $a_0 = \Delta z$. 
All the corresponding results collapses into a single curve by plotting the ratio $I(x,y)/I_0(x,y)$ as a function 
of  $\Delta z \times c_0$ and in a log-linear scale the variation is linear, as expected,  
for $a(x,y) c_0 < 0.15 g/m^2$ ($\mu = 2.9 \pm 0.4\ m^2/g$).
Using this result, and Eq. \ref{eq:bl}, the local, instantaneous concentration $c(x,y,t)$ during the displacement 
experiments can be determined from the following relation:
\begin{equation}
\frac{c(x,y,t)}{c_f} = \ln \left[ \frac{I(x,y,t)}{I_0(x,y)} \right]
\left/ \ln \left[ \frac{I_0(x,y)}{I_f(x,y)} \right] \right. .
\label{eq:aper}
\end{equation}
Figure~\ref{fig3}a displays a typical concentration map obtained in this way for an experiment in which the 
fracture walls are translated relative  to each other parallel to the arrow.
In addition to the irregularity of the front boundary, already discussed by \cite{AuradouHR01} 
and by \cite{DrazerAkH04}, its global shape is clearly anisotropic with a strong elongation in the direction 
perpendicular to the shear displacement ${\vec u}$. 
A similar anisotropy is observed on the aperture field displayed in Figure~\ref{fig3}b, 
with the appearance of elongated channels, also perpendicular to  ${\vec u}$.

\subsection{Quantitative characterization of anisotropy and relation to permeability field}
\label{subsec:sec3}

The anisotropy of the invaded zone can be characterized quantitatively from
 the inertia tensor  ${\mathbf M}(t)$ of the concentration distributions.
${\mathbf M}(t)$ is computed using the definition :
\begin{equation}
{\mathbf M}(t) = {1\over c_f}\sum c(x,y,t)
\left(
\begin{array}{cc}
y_*^2 & -x_* \times y_*\\
 -x_* \times y_* & x_*^2\\
\end{array}
\right),
\end{equation}
The sum is taken over all pixels in the image and $x_*=x-x_c$ and $y_*=y-y_c$ where $(x_c,y_c)$ 
is the  center of mass of the invasion pattern.
The eigenvalues  $\lambda_+^2$ and $\lambda_-^2$ of the tensor ${\mathbf M}(t)$ and the corresponding 
principal axes are then determined: they characterize the magnitude of the anisotropy and 
the directions of fastest and slowest flow, respectively.

More quantitatively, the tensor $M(t)$ may be used to estimate the ratio between the permeabilities for 
flow along these two directions. 
If we ignore the surface roughness and consider a two dimensional Hele-Shaw flow with 
{\em weak} anisotropy in the permeability,
$k_x\ne k_y$ but $|k_x-k_y| \ll (k_x+k_y)$, we can indeed compute the average
elliptical shape of a convected tracer front as follows.  Assume  liquid is
injected through a small hole of radius $r_0$ at pressure $p_0$, and allowed
to escape at the circular outer boundary $r=r_1$ at atmospheric pressure
$p_1\equiv 0$.  The velocities are
\begin{equation}
v_x = k_x {\partial p\over \partial x} \quad\mbox{and}\quad
v_y = k_y {\partial p\over \partial y},
\end{equation}
which together with the incompressibility condition $\nabla\cdot\vec{v}=0$
gives
\begin{equation}
k_x{\partial^2 p\over \partial x^2} + k_y {\partial^2 p\over \partial y^2} =0.
\end{equation}
To leading order in the permeability contrast, the pressure
is readily found to be
\begin{equation}
p(r,\theta) = {p_0\over\ln(r_0/r_1)}\left[ \ln(r/r_1)+{k_x-k_y\over 2(k_x+k_y)} \times 
({r^2-r_0^2-r_1^2+r_0^2r_1^2/r^2\over r_0^2+r_1^2}) \, \cos{2\theta}\,\right].
\label{eq:pr}
\end{equation}
Thus, to leading order in the permeability contrast, one has $v_x\sim k_x\ x/r^2$,
and a similar relation  for $v_y$.  The location of the right-most end of the tracer
ellipse, $(x_R,0)$, has then a velocity $dx_R/dt \sim k_x/x_R$ so that
$x_R^2(t) \sim k_x t$;  similarly, the coordinate of the top end of the tracer ellipse
verifies $y_T^2(t)\sim k_y t$. The ratio of the squares of the major and minor axes of the tracer 
ellipse, and therefore of the eigenvalues $\lambda_+^2$ and $\lambda_-^2$,  
is equal to the ratio of the permeabilities for flow parallel to these axis:
\begin{equation}
(\frac{\lambda_+}{\lambda_-})^2=\frac{k_{\perp}}{k_{\parallel}}
\label{eq:lambda}
\end{equation}
\section{Analysis of the experimental results}
\label{sec:sec3}
In this section we apply the procedures described above to  determine the permeability anisotropy 
from the shape of the invasion pattern during the radial injection of a dyed fluid that displaces 
a transparent one.
Several tracer injection experiments were performed for a fixed mean aperture $a_0=1\ mm$, and different
shear-displacements in both $x$ and $y$ directions. 
Reference test experiments with no shear displacement between the fracture walls were also performed  
to check that aperture variations are statistically isotropic 
(in this case, a roughly circular shape of the invaded zone is expected for a radial injection).

\begin{figure}
\includegraphics[width=\W]{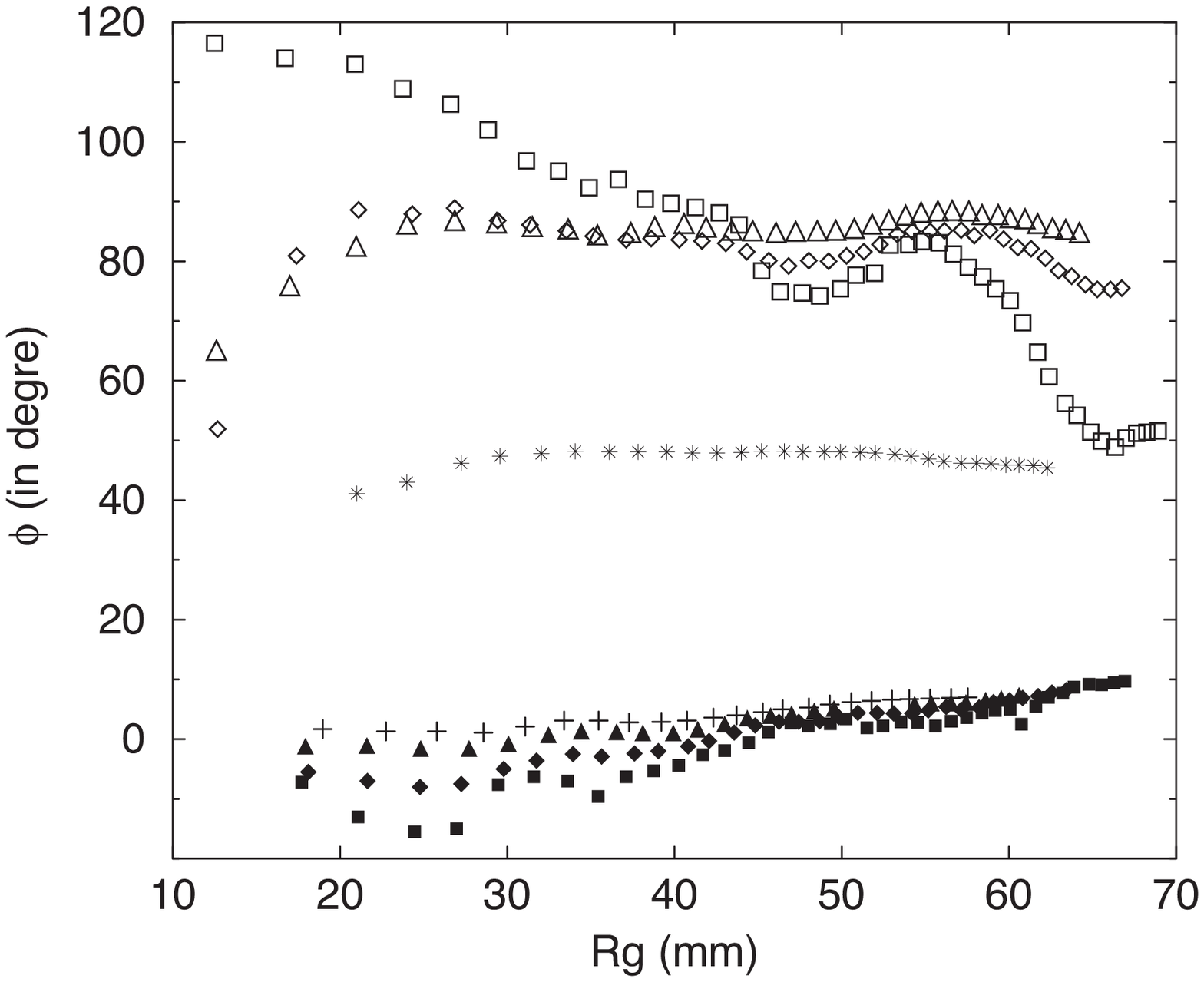}
\caption{Angle $\phi$ between the $x$ axis and the direction of maximum flow velocity
for different shear displacements. ${\vec u}$ parallel to $x$ :  u = 0.25 mm ($\square$), 0.5 mm ($\lozenge$), 0.75 mm ($\vartriangle$); $ {\vec u}$ parallel to $y$ :  u = 0.25 mm ($\blacksquare$), 0.5 mm ($\blacklozenge$), 0.75 mm ($\blacktriangle$), 1 mm ($+$); ${\vec u}$ at $-45^o$ from $x$, u = 1 mm ($\ast$).}
\label{fig4}
\end{figure}
The direction of the maximum flow velocity is associated to the principal axis of the inertia tensor 
${\mathbf M}(t)$ with the largest eigenvalue. Figure \ref{fig4} displays the variation of the angle 
$\phi$ of this axis with $x$ as a function of the radius of gyration $R_g$ of the invasion  pattern. 
$R_g$ verifies   $R^2_g = {\rm tr}({\mathbf M})$ and increases approximately as the square root of time 
as required by mass conservation.

The data in Fig~\ref{fig4} correspond to different magnitudes and orientations of the shear displacement ${\vec u}$
In the case of non-zero shear displacements oriented parallel to the $y$ axis (dark symbols), the angle $\phi$ 
remains close to zero, with an average over all data equal to $2.5^o$ and variations of the order of $\pm 5^{o}$ 
around that value. 
In addition, the curves corresponding to larger amplitudes of the shear displacement 
($|\vec{u}|=u_y$ varying from $0.5$ to $1$ mm) are all very similar. 
The elongation of the invaded zone is therefore, as expected, globally perpendicular to the lateral shift.

Analogous results are obtained for  displacements parallel to the $x$ axis (open symbols) and at
an angle of $-45^{o}$ from it ($\ast$ symbols).  
In these cases, $\phi$ is of the order of $90^{o}$ and $45^{o}$, respectively, indicating again 
that the invaded zone is elongated perpendicularly to the displacement.
This demonstrates that  the flow anisotropy is directly connected to the lateral shift,
with the preferential direction of flow oriented at an angle of $90^\circ$ 
with respect to the direction of the shear displacement:
this anisotropy  does not arise therefore from any intrinsic feature in the
fracture surface itself.

These results are in agreement with qualitative observations of the aperture fields, such as that presented 
in Fig. \ref{fig3}:  the shear displacement creates ridges in the aperture field oriented in the perpendicular
direction and partly inhibiting the flow.  
These conclusions are of course only valid if the shear displacement is different from zero. 
In the absence of such shear-displacement we observe, in some cases, a large dispersion in the 
measured values of $\phi$, as expected for a nearly circular invaded zone for which $\phi$ is undetermined. 
In other experiments, $\phi$ is much less dispersed, possibly due to a residual lateral shift.

\begin{figure}
\includegraphics[width=\W]{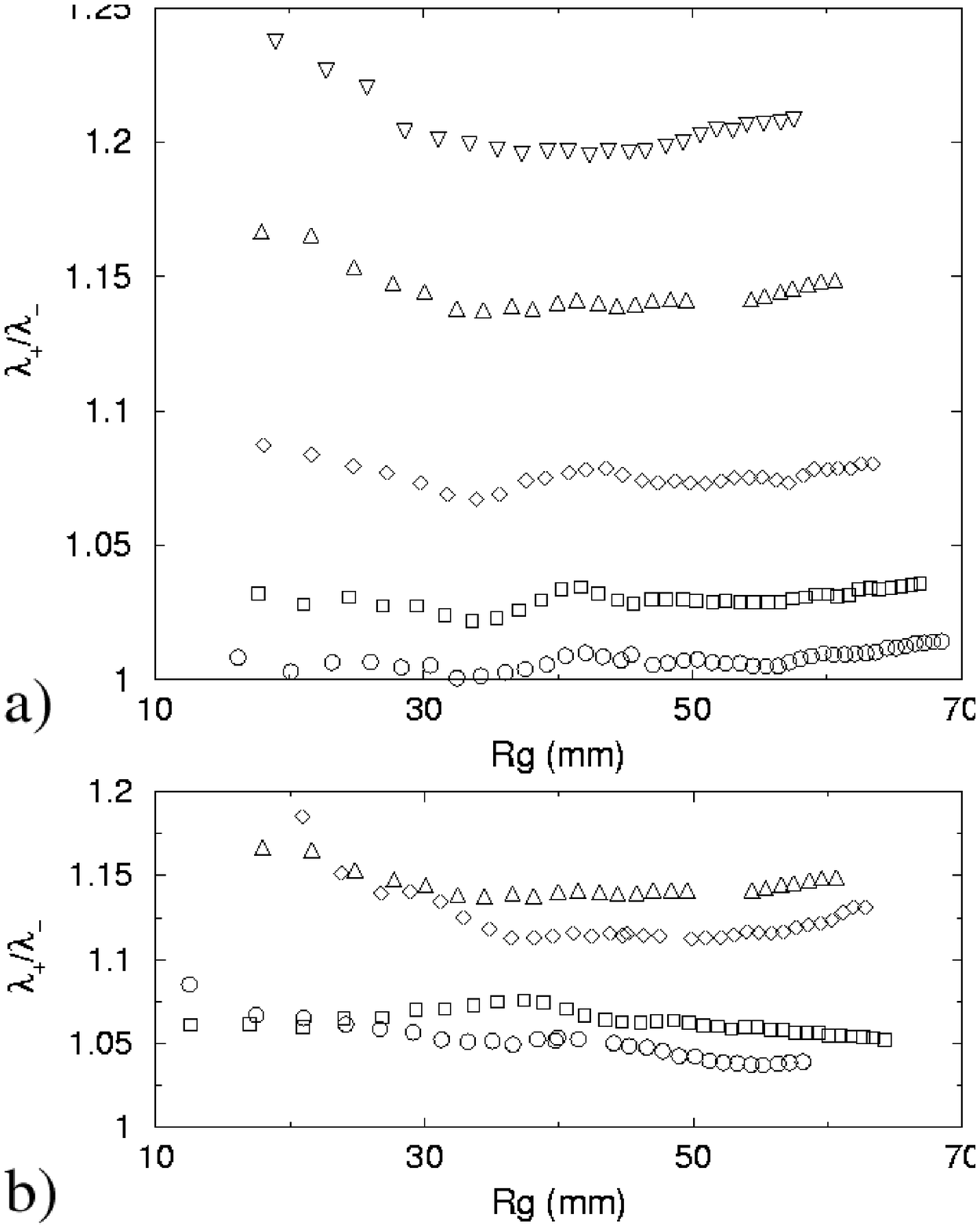}
\caption{Experimental variation of the ratio $\lambda_+/\lambda_-$ as function of the 
radius of gyration $R_g$ of the invaded zone for a shear displacement ${\vec u}$ oriented 
along $+y$ and with the following amplitudes : $u =$ 0 ($\circ$), 0.25 mm ($\square$), 
0.5 mm ($\lozenge$), 0.75 mm ($\triangle$), 1 mm ($\triangledown$).  
(b) Variation of  $\lambda_+/\lambda_-$ with $R_g$ for a constant magnitude of the shear displacement 
${\vec u}$ ($u = 0.75 mm$) and oriented in the following directions: 
+x($\circ$), -x ($\square$), +y ($\triangle$), -y ($\lozenge$).} \label{fig5}
\end{figure}
While $\phi$ indicates the orientation of the anisotropy, its magnitude is characterized by the ratio 
of the eigenvalues of the inertia tensor, $\lambda_+/\lambda_-$.
The variation of $\lambda_+/\lambda_-$ with the radius of gyration $R_g$ of the invaded zone is displayed 
in Figure~\ref{fig5} : $\lambda_+/\lambda_-$ reaches a roughly  constant value for
$R_g > 30\ mm$, suggesting that the growth of the invaded zone is self-similar.
This asymptotic value of
$\lambda_+/\lambda_-$  represents  then a robust parameter to characterize the anisotropy of the permeability 
for a given orientation and amplitude of the shear displacement.

Figure~\ref{fig5}b also demonstrates that, for a fixed amplitude of the shear displacement, its 
orientation may influence $\lambda_+/\lambda_-$. 
When the displacement is applied along a given orientation either in the positive or negative direction,
the effect on the ratio $\lambda_+/\lambda_-$ is weak. On the other hand, its magnitude is somehow
affected by changes in the orientation: $\lambda_+/\lambda_-  \simeq 1.05$ for 
$\vec{u}$ oriented along the $x$ axis, but it reaches $1.15$ when the shear displacement is applied in 
the $y$ direction. 
These fluctuations come from the statistical character of the distribution of the ridges that alter the flow: 
they vary not only from sample to sample, but also with the orientation of ${\vec u}$. 
On the other hand, for a given orientation, the variation of $\lambda_+/\lambda_-$ with the magnitude
of the shear displacement $|\vec{u}|=u$ is smooth, suggesting that  the location of the main ridges 
remains the same as $u$ increases. 
These results on  $\lambda_+/\lambda_-$ may be translated into permeability ratio by means of
equation~ \ref{eq:lambda}, which allows then for comparisons with the numerical simulations 
discussed below.

\section{Theoretical discussion of the relation between permeability anisotropy and aperture field}
\label{sec:sec4}
The self-affine nature of the fracture walls affects the void distribution
between them, and also plays an important role in fluid flow due to the
hydrodynamic boundary conditions imposed there.
Here we consider two complementary fracture surfaces with their mean planes
parallel and horizontal, which can be separated vertically in order to open
the fracture and shifted horizontally to introduce a shear displacement.
For a vertical separation $a_0$ and shear displacements $u_x$ and $u_y$ along
the $x$ and $y$ directions, respectively,
the local aperture is defined as the vertical distance separating the two
walls and is given by :
\begin{equation}
a(x,y) = h(x, y) - h(x-u_x,y-u_y) + a_0.
\label{eq:aper2}
\end{equation}

\begin{figure}
\includegraphics[width=\W]{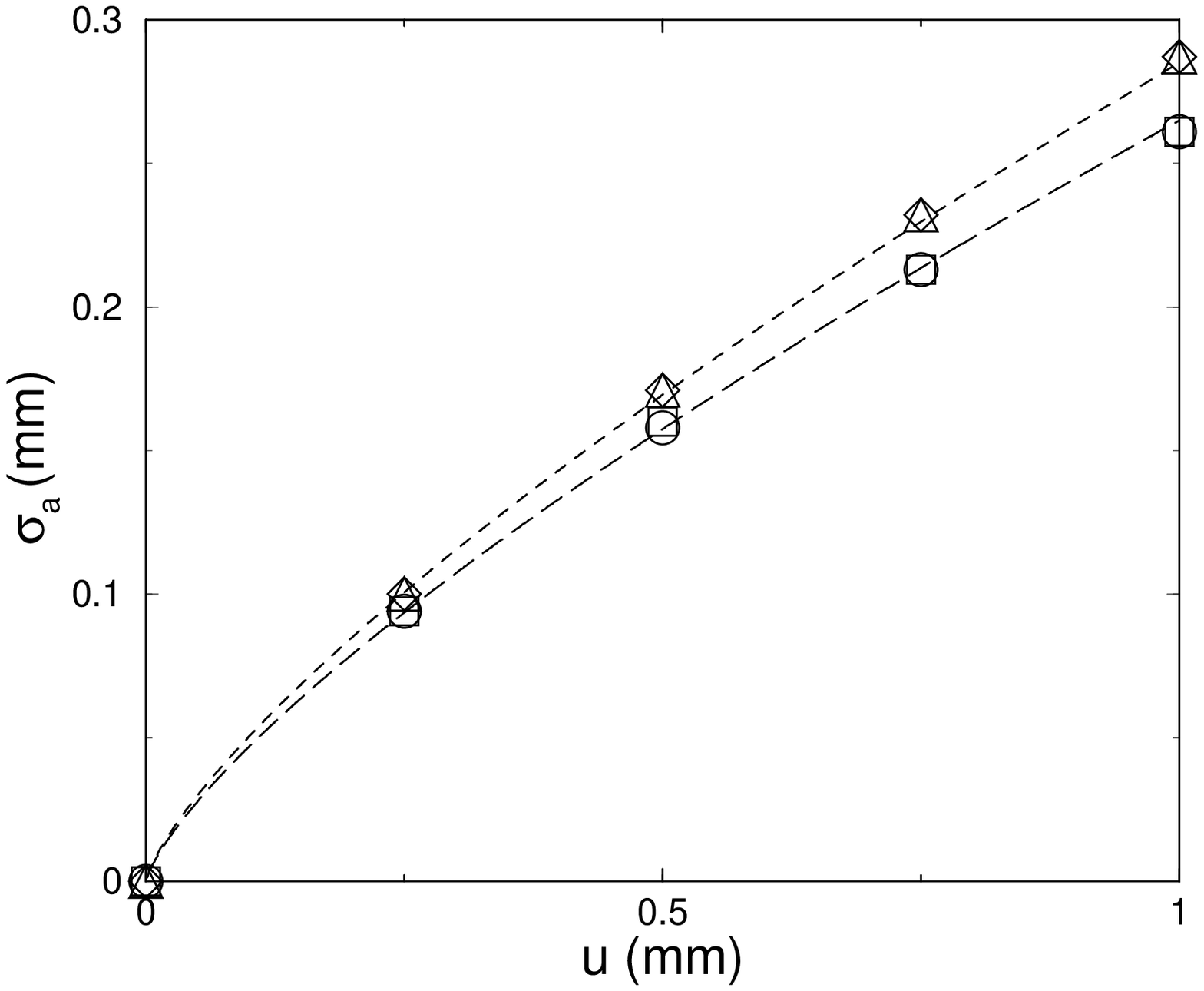}
\caption{Variance of the aperture field, $\sigma_a$, as a function of the magnitude of the 
shear displacement, $u$, in directions : +x($\square$), -x ($\circ$), +y ($\triangle$), -y ($\lozenge$). 
Dashed line $\sigma_a=0.285\ u^{0.75}$ and Long dashed line : $\sigma_a=0.265\ u^{0.75}$.}
\label{fig6}
\end{figure}

The covariance of the aperture was obtained previously (using Eq. \ref{eq:hh}), and two distinct 
scaling regimes were found \cite[]{PlourabouekHRS95}.
For distances smaller than the shear displacement, the aperture field was shown to have
self-affine correlations with exponent $\zeta$, while for larger distances, the
covariance decreases with the distance.
The magnitude of the shear displacement thus represents a crossover length scale, and can
be regarded as the typical size of features in the aperture distribution.
Considering the self-affine scale invariance, the aperture variance,
$\sigma_a^2=\langle (a(x,y) -  a_0)^2 \rangle$,
for a shear displacement of size $u$ is simply given by $\sigma_a(u) = C u^\zeta$, where $C$ is
related to the amplitude of the roughness of the fracture wall.
Figure \ref{fig6} demonstrates this power law behavior for different
orientations of the shear displacement directions. 
For all the chosen orientations $\sigma_a$ varies with $u$ with an exponent $\zeta=0.75$, 
in agreement with the Hurst exponent measured
in Sec. \ref{subsec:sec1} $(\zeta = 0.75 \pm 0.05)$.\\
In the following, we use
the ratio $S\equiv \sigma_a(u)/a$ as a parameter to quantify the
normalized fluctuations in the local aperture introduced by the shear displacement:
when $S<<1$, the aperture irregularities are a small perturbation on a
nearly-constant aperture field (note that in the absence of shear displacement between
the surfaces the aperture is constant and equal to $a_0$ throughout the fracture)
and their effect on the flow is negligible, but as $S$ approaches $1$ the channel becomes
quite convoluted and local aperture variations must be considered to correctly describe
the flow properties.

A second important earlier result is that the
aperture correlation length is anisotropic, being smaller in the shift
direction than in the orthogonal direction by a numerical factor of
$2\zeta-1 < 1$ \cite[]{RouxPH98}.
The effect of such an anisotropy on the flow in rough fractures was studied
numerically by \cite{ThompsonB91}, who
demonstrated that the  permeability
along the direction of the largest correlation length is enhanced while flow
parallel to the direction of the smallest correlation length is inhibited.
The effect was related to the occurrence of large ridges oriented along the
largest correlation length direction, i.e. in the direction
perpendicular to the shear displacement, and in fact such
large structures are observed in our system, as can be seen 
in Figure \ref{fig3}.

This description can be used to motivate a model for the permeability
anisotropy which considers the parallel ridges to provide parallel flow
channels, which are in series with respect to the direction of the shift
\cite[]{ZimmermanB96}.  Flow parallel to the shift is then analogous to
electrical current flow through resistors in series and the equivalent
hydraulic aperture is $\langle a^{-3}\rangle^{-1}$, whereas flow perpendicular
to the shift is analogous to resistors in parallel where the equivalent
aperture is $\langle a^3 \rangle$.  Specifically, assuming
that the distribution of apertures is Gaussian, one can evaluate
the two averages involved, yielding the upper and lower bounds
\begin{equation}
\frac{\langle a^3 \rangle}{a_0^3} = 1 + 3S^2,
\label{kupper}
\end{equation}
and
\begin{equation}
\frac{\langle a^{-3} \rangle^{-1}}{a_0^3} = 1-6S^2-9S^4+O[S^6],
\label{klower}
\end{equation}
respectively, in the limit $S<<1$.

Before applying this prediction to the experimental data, we shall first evaluate them on the results of simulations performed on numerical models of self-affine fractures. These models allow for an easier testing of the variability of the results from a fracture sample to another. They also allow to determine separately the permeabilities for flow parallel and perpendicular to the displacement vector ${\vec u}$, while the experiments only allow to determine the ratio of these two permeabilities.

\section{Numerical procedure}
\label{sec:sec5}
A first issue is the choice of the flow computation technique. When the local aperture of the 
fracture varies slowly along the fracture plane,
the lubrication, or Reynolds, approximation can be used to describe the flow in
the fracture \cite[]{MeheustS00}. On the other hand, many studies have pointed out that,
for natural fractures, this approximation fails to give an accurate description of the
hydraulic properties \cite[]{YeoDZ98,NichollRGD99,DijkB99,Konzukk04}.
In fact, when the roughness of the walls becomes non-negligible compared to the mean
aperture of the fracture, e.g. when the surfaces are close to contact, the lubrication
approximation fails, and a complete description of the flow field, which
accounts for the flow velocity normal to the surface, is needed \cite[]{MourzenkoTA95,Drazerk00}.
The lattice Boltzmann technique has therefore been selected to compute the flow because of its good 
adaptation to 3D complex geometries.

First, a self-affine surface is generated numerically, using a Fourier synthesis method \cite[]{Drazerk02},
and the fracture pore space is modeled as the region between one such surface and a suitably translated
replica. In order to mimic the experimental procedure, the two complementary surfaces are first assumed to 
be  separated by a fixed distance $a_0$ in the direction $z$ normal  to the mean plane of the fracture.  
Then, a shear displacement of the upper surface is introduced in the $(x,y)$ plane. The maximum shear
displacement investigated in each direction corresponds to the occurrence of the first contact point 
between the two surfaces and, therefore, the fracture remains open over its full area. 
In order to compare the numerical results to the experiments, both the Hurst exponent and the average 
magnitude of the surface roughness need to be matched to their experimental values: 
the average fluctuations in surface height over a distance of $1\ \rm{mm}$ was thus been set to 
$0.25\ \rm{mm}$, and the Hurst exponent $\zeta$ was set to $0.8$. 
Finally, the mean aperture was set to $a\simeq1$mm, which is comparable to that used in the experiments.  
Using these values, the local height varies considerably over distances of the order of the aperture 
size so that computing the complete flow field  inside the fractures is required.

In order to capture the three-dimensional character of the flow field inside the fractures large cubic
lattices, with as many as $1024 \times 1024 \times 20$ sites, are used in the simulations. Hence, the
lattice spacing $\delta$ corresponds to   $0.05\ \rm{mm}$ (the mean aperture $a_0=1$mm corresponds
to $20$ lattice sites). The Lattice Boltzmann method is then used to compute the flow field in the presence
of a pressure gradient and using periodic boundary conditions in the plane of the surface
(see \cite{Drazerk00} for details of the simulation method). The pressure gradient is applied both
along and perpendicular to the shear displacement between surfaces. In addition, and to avoid the possible
effects of any intrinsic anisotropy in the generation of the self-affine surfaces, the shear displacements
are introduced in both the $x$ and $y$ directions within the mean plane. 
The Reynolds number in the simulations is $\rm{Re} = Ua/\nu \sim a^3 \rho \nabla P / 12 \nu^2 < 1$ 
for $a=1$mm, where we used the parallel plane approximation to estimate $U$. Inertia effects are 
therefore negligible \cite[]{SkjetneHG99}, and the flow is governed by the linear Stokes
equations, which are invariant under velocity rescaling. Then, by computing the flow rate, we can 
obtain the permeability of the fracture and its dependence on the magnitude and orientation of the 
shear displacement.

\section{Analysis of the numerical simulations}
\label{sec:sec6}
Numerical simulations are performed in configurations reproducing as far as possible those encountered 
in the experiments. Several series of computations were performed on   self-affine fracture models 
for different mean apertures ($0.5<a_0<2.0\,mm$) and shear-displacements ${\vec u}$ ($0<u<1.6\,mm$). 
The displacement ${\vec u}$ is oriented along directions $x$ or $y$ in the mean plane
of the surfaces and the size of the simulated fractures is typically $25.6 \times 25.6$\,mm. 
For given values of $a_0$ and ${\vec u}$, computations are performed on 10 independent realizations 
in order to estimate the dispersion of the results (only a part of the corresponding data will be 
displayed for clarity but all the data is consistent with the trends visible on the figures).
\begin{figure}
\includegraphics*[width=\W]{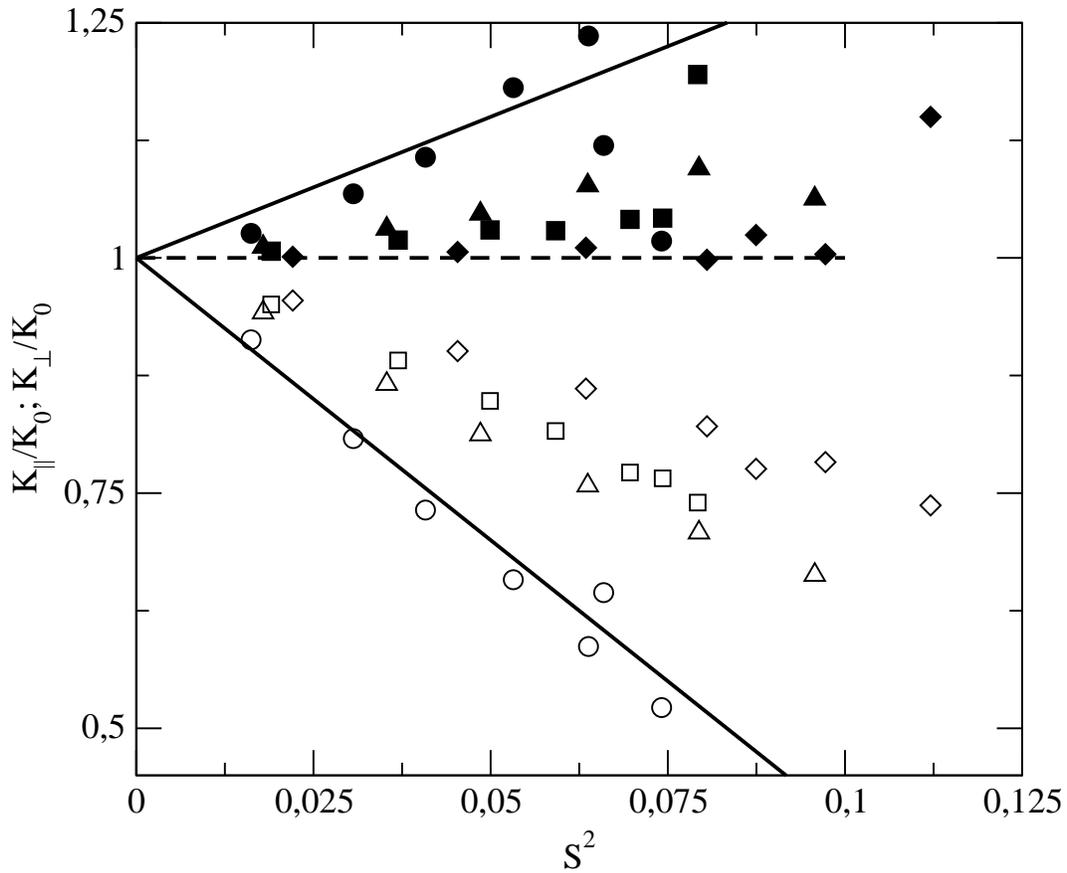}
\caption{Permeability as a function of the shear displacement in self-affine fractures.
Open (closed) symbols correspond to the permeability parallel (perpendicular) to the 
shear displacement $k_{\parallel}/k_0$ ($k_{\perp}/k_0$) as a function of the normalized 
aperture variance $S^2$. 
The symbols [($\scriptstyle{\blacksquare,\square}$) and ($\bullet,\circ$)]
correspond to simulations performed in a single fracture but for two orthogonal 
directions for the shear displacement, that is $u_x \neq 0$, $u_y = 0$ and $u_x = 0$, $u_y \neq  0$, 
respectively. Analogously, the other four sets, 
[($\blacklozenge,\lozenge$) and ($\blacktriangle,\triangle$)] correspond to simulations performed 
in a single fracture and for two orthogonal orientations of the shear displacement.
The upper (lower) solid line corresponds to the upper (lower) bound of the permeability given
in Eq.~\ref{kupper} (Eq.~\ref{klower}).} \label{fig7}
\end{figure}

Figure \ref{fig7} displays the variation of the normalized permeabilities  
$k_{\parallel}/k_0$ (resp. $k_{\perp}/k_0$) as a function of the normalized aperture variance $S^2$ for a 
constant mean aperture $a_0 = 1.0\,mm$ and for displacements either along $x$ or along $y$.  $k_0$ is the 
isotropic permeability for $u=0$ and $k_{\parallel}$ and $k_\perp$ are the respective permeabilities  for 
flow parallel and normal to the shear displacement. The normalized variance $S^2$ is used as the horizontal 
scale since, as already discussed in Sec.~\ref{sec:sec4}, it is the relevant parameter to characterize the 
dependence  of the permeability  on the shear-displacement (the magnitude $u$ of the shear displacement 
ranges between  $0$ and $1.6$\,mm).

The key feature is the fact that $k_{\perp}/k_0$ and $k_{\parallel}/k_0$ both vary linearly with $S^2$ but 
in opposite directions ($k_{\perp}/k_0$ increases with $S^2$ while $k_{\parallel}/k_0$ decreases). 
Figure~\ref{fig7} also shows that this linear variation is observed for shear displacements $\vec u$ 
oriented both parallel to $x$ and $y$ axes, which demonstrates  that the permeability anisotropy is 
induced by the shear displacement and it is not associated to geometrical anisotropies present before 
the relative displacement of the walls.

These linear variations of the permeability qualitatively agree with the 
simple model presented in Sec.~\ref{sec:sec4}. In fact, in Sec.~\ref{sec:sec4}
we estimated the upper bounds for the variations of $k_{\perp}/k_0$ and $k_{\parallel}/k_0$ 
with $S^2$ (see Eqs.~\ref{kupper} and \ref{klower}), which are represented by solid lines in 
Fig.~\ref{fig7}. As can be seen, the sign and order of magnitude of these variations are the 
same as those of the numerical simulations.

Another interesting result is the strong correlation between the variations of $k_{\parallel}/k_0$ 
and those of $k_{\perp}/k_0$ : fractures displaying a large reduction of the permeability 
$k_\parallel$ are generally the same that display a significant  increase of  $k_\perp$.

\begin{figure}
\includegraphics*[width=\W]{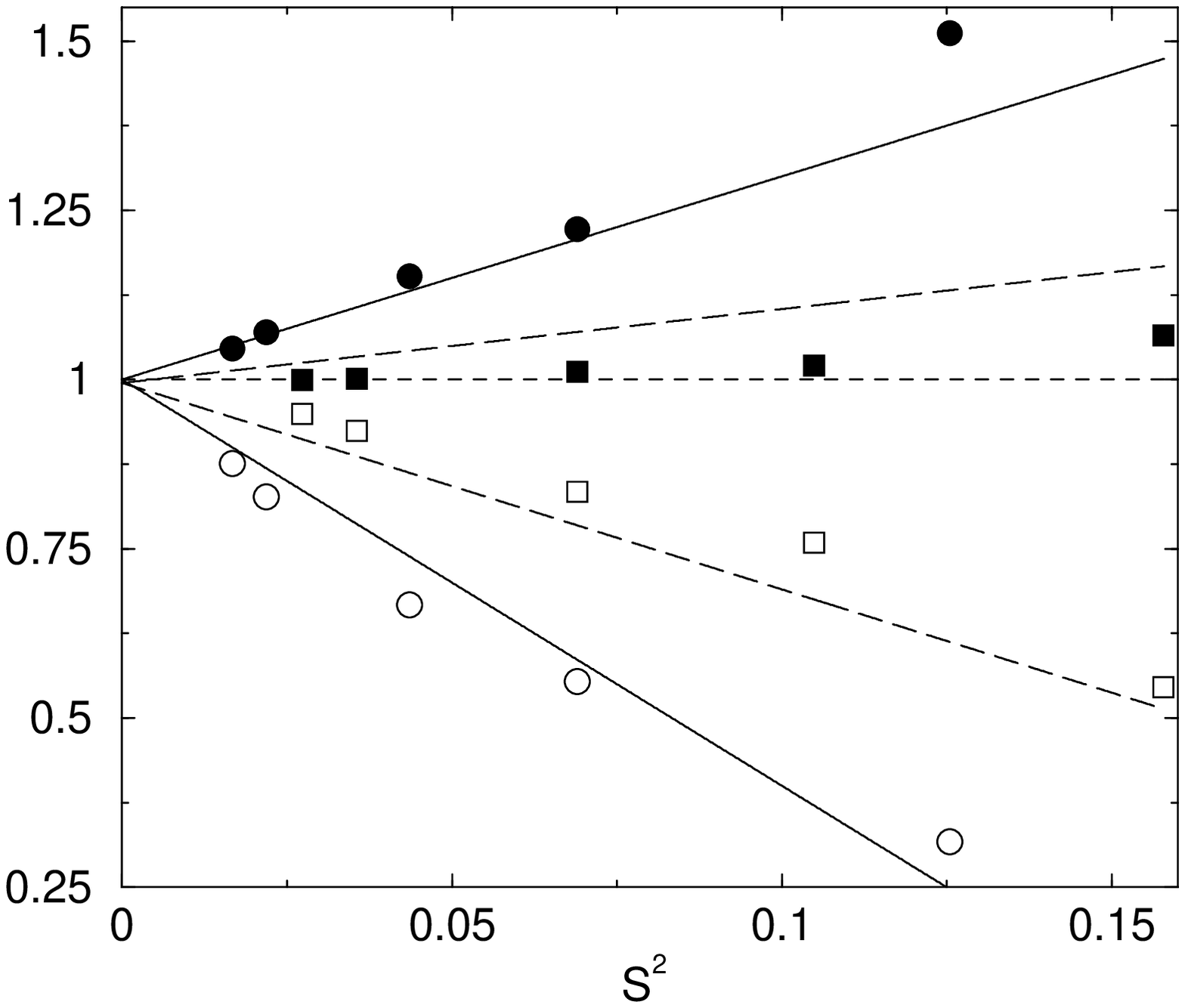}
\caption{Normalized permeabilities $k_{\parallel}/k_0$ (open symbols) and $k_{\perp}/k_0$ (closed symbols) 
from two numerical simulations as a function of the normalized aperture variance $S^2$. 
Simulations correspond to a fixed shear displacement $u_x = 1 mm$ and variable aperture 
$0.75 < a_0 < 2.0$\,mm. The upper (lower) solid line correspond to the upper (lower) 
bound for the permeability given in  Eq.~\ref{kupper} (Eq.~\ref{klower}).
The long dashed lines correspond to a linear fit of $k_{\perp}/k_0$ and $k_{\parallel}/k_0$ 
over all the numerical realizations.} \label{fig8}
\end{figure}

Finally, other simulations were performed in which $S^2$ is modified by varying the mean aperture 
from $a_0 = 2.0$\,mm to  $a_0 = 0.75$\,mm, while keeping the shear displacement constant, $u = 1.5$\,mm. 
In this case, the variance $S^2$ increases when the aperture $a_0$ decreases.  
The variations of $k_{\parallel}/k_0$ and $k_{\perp}/k_0$ as a function of $S^2$ for sets of data obtained 
in this way for three different realizations of the fracture surface are plotted in Fig.~\ref{fig8}. 
One observes the same general trend as when $a_0$ is  kept constant and $u$ varies like in 
Fig.~\ref{fig7} : $k_{\parallel}/k_0$ decreases  and $k_{\perp}/k_0$ increases or remains 
constant as $a_0$ is reduced and $S^2$ increases.  For a given realization of the surface, the 
variations of $k_{\parallel}/k_0$ and $k_{\perp}/k_0$ with $S^2$ are linear and both remain within the 
bounds set by Eqs.~\ref{kupper} and  \ref{klower}. 
This confirms that the normalized aperture variance, $S^2$, is indeed the adequate parameter to 
characterize the disorder in the aperture field introduced by the shear displacement between 
the two surfaces.

Numerically, $10$ self-affine surfaces were generated and were used to create aperture fields for 
different mean apertures and shear displacements. 
In average, fitting the permeability values to a quadratic dependence on $S$ over all the 
realizations we obtain:
\begin{equation}
\left\langle \frac{k_{\perp}}{k_0} \right\rangle = 0.995 + 1.09 S^2,
\end{equation}
and
\begin{equation}
\left\langle \frac{k_{\parallel}}{k_0} \right\rangle = 0.955 - 3.05 S^2.
\end{equation}
Nevertheless, large fluctuations are observed from one realization to another: 
for some the wall roughness only weakly perturbs the flow while for some others the roughness 
strongly distorts the flow and the permeability behaves as 
predicted by Eqs.~\ref{kupper} and \ref{klower}.

\section{Comparison between experiments and simulations}
\label{sec:sec7}

\begin{figure}
\includegraphics*[width=\W]{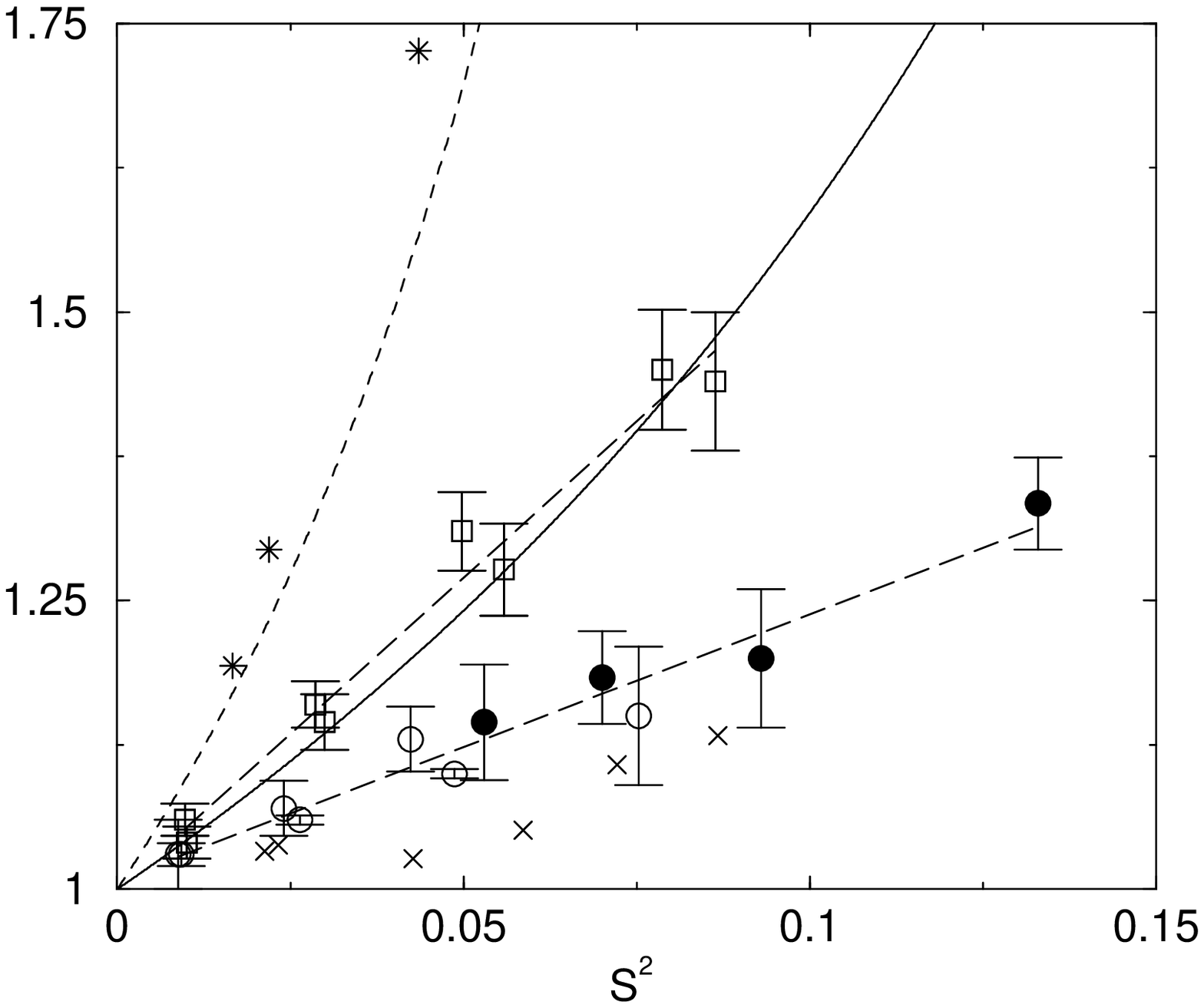}
\caption{
Permeability ratio $k_{\perp}/k_{\parallel}$ as a function of the normalized aperture variance $S^2$.
($\square,\circ,\bullet$) experimental data; ($\star$, $\times$) numerical data.
$\square$ symbols correspond to a shear displacement along the $y$ axis
(either in the positive or negative direction).
$\circ$ symbols correspond to a shear displacement along the x-direction with $a_0=1.0\ mm$, 
while $\bullet$ symbols correspond to $u_x=2.0\ mm$ and mean apertures ranging from $1.25\ mm$ to $2\ mm$.
$\star$ and $\times$ symbols correspond to numerical realizations showing the strongest and the 
weakest anisotropy, respectively.
The dotted line corresponds to the upper bound for the permeability ratio estimated from 
Eqs.~\ref{kupper} and \ref{klower}. 
The solid line corresponds to the permeability ratio $\langle k_{\perp}/k_{\parallel} \rangle$ 
averaged over all the numerical realizations.
The long dotted lines corresponds to linear fits of the experimental data.
}
\label{fig9}
\end{figure}


As discussed in Sec.~\ref{subsec:sec3},  the analysis of the geometry of the invasion pattern in the 
injection experiments allows  one to estimate the ratio $k_{\perp}/k_{\parallel}$ of the permeabilities
for flow perpendicular and parallel to the shear displacement ${\vec u}$. The variance $S^2$ of the 
aperture can be computed for each set of values of $a_0$ and ${\vec u}$ from the profilometer maps 
of the fracture surfaces (see Sec.~\ref{subsec:sec1}).  It is therefore possible to determine 
experimentally both $k_{\perp}/k_{\parallel}$ and $S^2$ for each experiment.
Figure \ref{fig9} displays the variations of the permeability ratio as a function of $S^2$, as 
determined from the shape of the invasion front in a series of injection experiments 
(see Sec.~\ref{subsec:sec3}).
Three sets of experiments are presented, two of them correspond to shear displacements oriented 
along the x-direction or y-direction.
No distinction between the positive or negative direction is made since, as point out in Sec.~\ref{sec:sec3}, 
the permeability ratio is roughly independent of the shear displacement sign, but only on its magnitude 
and orientation. The third set represent experiments performed with a fixed shear displacement, 
$u_x=2.0$\,mm, and a variable aperture $a_0$ ranging from $1.25\ mm$ to $2.0\ mm$.
As it can be seen in Fig.~\ref{fig9}, the permeability ratios measured with a variable aperture and a 
shear displacement oriented along the $x$-direction are aligned with those obtained with a fixed 
aperture and a variable shear displacement applied along the x-direction.

Independently of the shear-displacement direction the permeability ratio increases roughly 
linearly with $S^2$ and its values remain bellow the upper bond provided by combining 
Eqs.~\ref{kupper} and \ref{klower}. We also show that the permeability ratios obtained from the 
experiments always lie between the strongest and weakest anisotropies observed in the very extensive
numerical computations. 
When $S \sim 0.1$, a departure of $k_{\perp}/k_{\parallel}$ from a linear behavior is observed 
in the numerical simulations. This result is consistent with the model discussed
in Sec.~\ref{sec:sec4}, which is clearly no longer valid for $S \gtrsim 0.13$, 
for which it predicts a closed fracture (zero permeability).
A sharp decreases of the permeability along the shear displacement is thus expected 
and corresponds to a strong increases of $k_{\perp}/k_{\parallel}$ when $S$ can no longer be considered 
to be small compared to $1$. 
This divergence is not observed experimentally since the method used for the anisotropy characterization 
is strictly valid for weak anisotropic permeabilities as discussed in Sec.~\ref{subsec:sec3}.

Finally, we also present in Fig.\ref{fig9} a linear fit to the experimental data.
Circumstantially, the strongest effect is observed when the walls are shift in the y-direction and a linear 
regression indicates an increased of $k_{\perp}/k_{\parallel}$ with $S^2$ with a slope of $5.4 \pm 0.3$. 
In the x-direction the permeability ratio also increased with $S^2$ but with a slope of $2.3 \pm 0.15$.

\section{Conclusions}
\label{sec:sec8}
The present study has confirmed both experimentally and numerically that the permeability of a fracture 
made of two complementary surfaces with a shear displacement $\vec u$ is lower for flow parallel to $\vec{u}$
than perpendicular to it.  
More precisely, the parallel permeability ($k_\parallel$) decreases significantly while the 
perpendicular one ($k_\perp$) increases or remains constant.

Qualitatively, this result appears to be related to the appearance of preferential channels of larger local 
aperture oriented perpendicular to $\vec u$: these channels are indeed visible on aperture images 
reconstructed from experimental (or numerical) profilometry maps.  Assuming such a geometrical structure 
allows to develop simple theoretical predictions for the relation between the permeability variations and 
the normalized variance $S^2$ of the aperture field. The permeability for flow parallel (perpendicular) 
to $\vec u$ was then shown to decrease (increase) approximately linearly with $S^2$. 
The absolute magnitude of the variation is also predicted to be two times larger for the parallel flow.  
These predictions are qualitatively correct and provide an estimation of the order of magnitude for the 
permeability variations and for the ratio of the permeabilities perpendicular and parallel to the 
shear displacement. In particular, both for the experiments and the numerical simulations, a linear 
variation for the permeability ratio with $S^2$ is actually observed - showing that the variance $S^2$ 
is a key parameter of the process. However there is a significant dispersion in the absolute value of the 
prefactors in these linear relations from one sample to another.

Further studies will be needed to understand the origin of the dispersion of the numerical values of the 
permeability variations and uncover the other characteristic geometrical parameters which may be involved 
in the process (finite size effects may  be a limitation,  particularly for the present simulations). 
Another important issue for practical applications is whether an estimation of the aperture variance can 
be obtained, for a given shift $\vec u$ and mean aperture $a_0$ of the fracture without requiring to map 
the full surface of the sample.

\begin{acknowledgments}
We are indebted to S. Bourles for providing us with the fracture moldings and
to G. Chauvin, R. Pidoux and C. Saurine for their assistance in the realization of the experimental
set-up.  Computer resources were provided by the National Energy Research
Scientific Computing Center.
HA and JPH are supported by the CNRS and ANDRA through the GdR
FORPRO (contribution No. $2004/20$) and the EHDRA (European Hot Dry Rock Association)  and PNRH programs.
GD and Jk are supported by the Geosciences Program of the Office of Basic
Energy Sciences (US Department of Energy), and by a PSC-CUNY grant.
This work was facilitated by a CNRS-NSF Collaborative Research Grant.
\end{acknowledgments}

\bibliography{articles,book,mios,mybooks}
\bibliographystyle{agu04}

\end{article}
\newpage

\end{document}